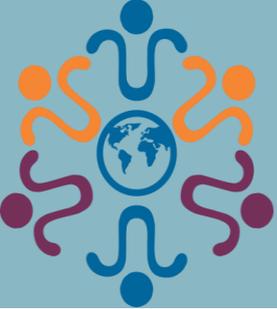

# MastitisApp: a software for preventive diagnosis of mastitis in dairy cows

# MastiteApp: um software para diagnóstico preventive de mastite em vacas leiteiras

# MastitisApp: un software para el diagnóstico preventivo de mastitis en vacas lecheras

**Italo Henrique Souza Mafra**
Graduated in Technology in Systems Analysis and Development
Institution: Instituto Federal de Educação, Ciência e Tecnologia de São Paulo (IFSP)
Address: Campinas – São Paulo, Brazil
E-mail: italohsmafra@hotmail.com

**Glauber da Rocha Balthazar**
PhD in Agricultural Systems Engineering
Institution: Instituto Federal de Educação, Ciência e Tecnologia de São Paulo (IFSP)
Address: Campinas – São Paulo, Brazil
E-mail: glauber.balthazar@ifsp.edu.br
Orcid: https://orcid.org/0000-0002-1993-6621

**ABSTRACT**
Dairy farming has great economic value in Brazil, however, during production, diseases such as mastitis can occur in animals, which can reduce productivity and, consequently, economic profitability. When mastitis is present in animals, it can cause physical and chemical changes in the milk, affecting its quality, market value and also compromising the health of the animal. MastiteApp is a tool to help producers prevent mastitis in their herds by checking the temperature taken from the four teats of the animal. To perform the analysis, the temperature of all the animals' teats must be measured and, if there is a change in temperature, the system will display a message informing the producer of the possible presence of subclinical mastitis in their animal. The application has proven to be efficient in alerting producers to the possible presence of subclinical mastitis in the first few days of manifestation, thus initiating treatment and preventing the disease from worsening.

**RESUMO**
A pecuária leiteira possui um grande valor econômico no Brasil, entretanto durante a produção podem ocorrer doenças no animal, tal como a mastite, que podem diminuir a produtividade e, consequentemente, a rentabilidade econômica. A mastite quando presente no animal pode causar alterações físico-químicas no leite afetando sua qualidade, o valor no mercado e também comprometendo a saúde do animal. O MastiteApp vem como uma ferramenta para auxiliar o produtor na prevenção da mastite em seu rebanho através da averiguação da temperatura extraída dos quatro tetos do animal. Para a realização da análise deve-se medir a temperatura de todos os tetos dos animais e caso apresente uma alteração na temperatura o sistema exibirá uma mensagem informando ao produtor uma possível presença de mastite subclínica em seu animal. A aplicação se mostrou eficiente para o fim de alertar o produtor com a possível presença de mastite subclínica nos primeiros dias de manifestação, iniciando assim o tratamento evitando que a doença se agrave.

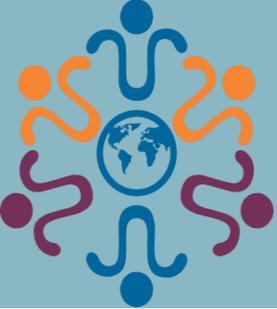

**Palavras-chave:** mastite, termografia, aplicativo android, pecuária.

**RESUMEN**
La producción lechera tiene gran valor económico en Brasil, sin embargo, durante la producción, pueden ocurrir enfermedades en el animal, como mastitis, que pueden reducir la productividad y, en consecuencia, la rentabilidad económica. La mastitis, cuando está presente en el animal, puede provocar cambios físico-químicos en la leche, afectando su calidad, valor de mercado y comprometiendo también la salud del animal. MastiteApp surge como una herramienta para ayudar a los productores a prevenir mastitis en su rebaño mediante el control de la temperatura tomada de los cuatro pezones del animal. Para realizar el análisis se debe medir la temperatura de todos los pezones de los animales y si hay un cambio de temperatura, el sistema mostrará un mensaje informando al productor de la posible presencia de mastitis subclínica en su animal. La aplicación demostró ser eficiente para alertar al productor sobre la posible presencia de mastitis subclínica en los primeros días de manifestación, iniciando así el tratamiento, evitando que la enfermedad empeore.

**Palabras clave:** mastitis, termografía, aplicación android, cría de vacas.

# 1 INTRODUCTION

In the second quarter of 2023, in Brazil, the acquisition of raw milk by establishments operating under some sanitary inspection was 35.40 billion liters (Brasil, 2024). According to the Brazilian Institute of Geography and Statistics – IBGE (IBGE, 2024), these are significant numbers when it comes to the Brazilian agricultural sector. However, during production, producers are faced with several diseases related to milk. One of the main diseases that affect dairy cattle is mastitis, which according to Nogueira *et al.* (2018), is the result of the immune response to some aggression suffered by the mammary tissue, with immediate consequences for the animal, as it causes pain and discomfort in the affected cattle. Now, the most studied forms of contamination are of microbial origin. Mastitis is classified into two types that are grouped according to their form of manifestation: clinical or subclinical. It is called clinical mastitis when signs of inflammation are evident and subclinical mastitis when the inflammatory process requires field tests such as the California Mastitis Test (CMT) or laboratory tests such as the Somatic Cell Count (SCC) for diagnosis. Field tests are simpler and can be performed daily at the time of milking (Ribeiro Júnior *et al.,* 2008).

According to Rezende (2017), the main losses caused by bovine mastitis include reduced milk production due to subclinical mastitis, increased production costs due to cases of clinical mastitis, increased disposal costs and premature death of some animals, increased expenses with veterinary fees and purchase of medicines, disposal of milk with antimicrobial residues, among others. Mastitis can also pose a health risk to milk consumers, mainly due to the presence of bacterial agents eliminated through milk. Another potential risk factor is the possibility of antibiotic residues in milk (Silanikove *et al.,* 2010).

Based on a study conducted on cows with mastitis, losses were calculated that could range from R$ 1,333.90 to R$ 2,145.89 per infected cow and from R$ 0.2146 to R$ 0.4311/kg of milk, not including treatment costs during the period in which the cow is infected (Lopes *et al.,* 2011).

It can therefore be concluded that, given all the problems caused by mastitis to producers, the need for prevention and early detection is essential, both to reduce losses from infected cows and for the well-being of the animal.

Therefore, due to the great economic importance of dairy farming in Brazil and the difficulty that producers have in detecting subclinical mastitis, it would be remarkably useful for producers to use technology to support them in detecting mastitis. Some research is being conducted using thermographic cameras and heat frequency image analysis. Thus, some researchers have already used thermography to detect mastitis in cattle (Rezende, 2017; Bortolami *et al.,* 2015) and obtained satisfactory results in detecting subclinical mastitis even before the producer noticed any visual changes in the infected cow. It is understood that the temperature of the animal's teats undergoes significant changes when diagnosed with mastitis. However, the hardware and software resources for thermography are expensive, which makes this detection expensive. Thus, it is clear that there is a fertile field of research for the production of technologies that analyze teat temperature to detect mastitis, but with the aim of reducing the cost of the hardware and software technologies involved. The main motivation of this study is to build a technology that is easily accessible to rural producers, aiming to inform them that a cow is suspected of having subclinical mastitis.

**2 THEORETICAL FRAMEWORK**

2.1 MASTITIS

Mastitis is the inflammation of the mammary gland parenchyma, regardless of the cause, characterized by a series of physical and chemical changes in the milk as well as pathological changes in the glandular tissue (Benedette *et al.,* 2008).

Mastitis, or mammitis as it can be called, is an inflammation of the cow's teat, caused by microorganisms such as bacteria and fungi as the main causes. Its main means of contagion occur during milking of animals through indirect contact between a sick cow and a healthy cow, such as through teat cups or the milker's hands. The cow can also be contaminated through the environment in which it lives, especially where there is an accumulation of manure, urine, mud and organic bedding. The main consequences in an infected cow can range from physical changes in the teat and udder to chemical-physical changes in the milk. (Rezende, 2017). Mastitis can be classified into two distinct levels: clinical and subclinical, both of which are the same disease, but at distinct stages of evolution. The symptoms of subclinical mastitis begin to appear more explicitly for the producer when the mastitis has already evolved into clinical mastitis, where the main symptoms in cows are an increase in the volume of the teat, an increase in the temperature of the udder and hardening of the mammary gland. To confirm the presence of the disease in the cow, it is advisable to perform the Sieve Test, which aims to identify granules in the milk in a cup with a dark background (Ruegg, 2012). Subclinical mastitis does not present visible symptoms in the udder and milk as in clinical mastitis, but there are some known tests that can be used to detect subclinical mastitis. The way to find out whether the cow is contaminated or not is through the Somatic Cell Count (SCC). The main tests that can be performed to detect mastitis are the California Mastitis Test (CMT) and the Wisconsin Mastitis Test (WMT), which seek to detect an increase in SCC in the milk. During the inflammatory period of the cow, the cow's immune system produces defense cells to try to reverse the inflammatory process that has begun in the cow. These defense cells are called somatic cells and, when contaminated, the somatic cell count is generally above 300,000 cells/ml of milk (Santos; Fonseca, 2007).

## 2.2 THERMOGRAPHY

Infrared thermography is a technique for capturing and analyzing thermal information from devices with the specific purpose of capturing heat without direct contact with the animal or object. Infrared is an electromagnetic frequency naturally emitted by any animate or non-animate object that is capable of producing, reflecting or absorbing heat, with an intensity proportional to its temperature (Leão *et al.,* 2015).

Thermography originated with the Greek physician Hippocrates, the famous "Father of Medicine", who applied mud to the body of the sick person and then analyzed the areas where the dried mud was hotter than the rest, therefore this was the point where the disease was located (Brioschi, 2003).

Infrared imaging equipment is capable of detecting temperature with a variation in temperature of 0.05 °C, while the human hand is not capable of perceiving temperatures lower than 2 °C to 4 °C. The use of thermography in veterinary medicine is of considerable importance due to the difficulty in communication between the patient (animal) and the doctor (Rezende, 2017). The temperature on the surface of animals can change depending on the blood flow in the region measured by the sensor. Mastitis, as an infection, leads to an inflammatory process in the infected region and, in response, the immune system alters the blood flow in the affected region and, consequently, the region undergoes a temperature variation (Leão *et al.,* 2015). Some studies conducted by professionals in the area of veterinary medicine have proven through their studies that the use of infrared data analysis is effective in detecting subclinical mastitis, since their analyses were able to detect changes in the temperature of infected mammary glands in those that were healthy. To validate the infrared tests in animals with mastitis, control tests such as CMT were performed, which served to prove the effectiveness of their methods (Rezende, 2017; Nogueira, 2013).

## 2.3 ANDROID

Android is a software environment written for mobile devices. The Android environment is mainly composed of the Operating System (OS) developed on a Linux kernel, a User Interface (UI), user applications (WhatsApp, phone, Facebook, etc.), code libraries, application frameworks, among others (Ableson, 2012). According to the International Data Corporation (IDC, 2020), the Android system is the most widely used in the world and its share of use in smartphones has been increasing eremarkably year, as can be seen in Table 1.

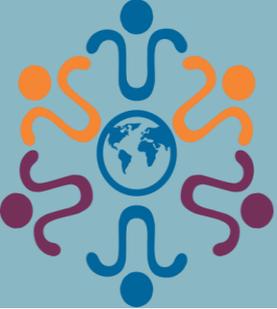

Table 1. Market share of mobile devices in the world

| Year | 2017 | 2018 | 2019 | 2020 |
|---------|-------|-------|-------|-------|
| Android | 85.1% | 85.1% | 86.6% | 86.6% |
| iOS | 14.7% | 14.9% | 13.4% | 13.4% |
| Others | 0.2% | 0.0% | 0.0% | 0.0% |

Source: IDC (2020)

Android is built on a Linux kernel and a virtual machine (VM) optimized to run Java applications. Google chose Java due to the number of Java software developers and the rich environment that Java programming has to offer (Ableson, 2012).

The fact that Android is the most widely used OS for smartphones and also uses the Java language to develop applications (apps) were fundamental requirements for choosing the environment and language for the development of this project, given that the disciplines of the Technology in Systems Analysis and Development course deal specifically with these technologies.

## 2.4 DATABASE

For data persistence, a Relational Database (RDB) was used in this work, that is, one that follows the entity relationship standard for data storage where the stored values must be saved in rows and columns. For this purpose, the MySQL database was chosen due to it presents the relational characteristics necessary for the project and due to it is free and freely usable. Thus, the data is stored in a MySQL database, which is a database management system that uses the Structured Query Language (SQL) as an interface (Prates, 2000). Some of the main positive points for its use are that it can be executed on a server, has good performance, is reliable, easy to use and free (Pina, 2015).

2.5 SCRUM

In 2001, a meeting of 17 leaders working in the software industry discussed new ways of working in order to develop a new software production methodology that could be used by all software developers. The group reached a consensus that some principles were essential for achieving good results. As a result, the twelve principles and the publication of the Agile Manifesto emerged (Fadel, 2010).

Initially, Scrum was designed for managing automotive and consumer product manufacturing projects by Takeuchi and Nonaka in 1986 at Harvard University. They found that projects with small, multidisciplinary teams produced more efficient results. In 1995, Scrum was redefined by Ken Schwaber to consolidate it as a software development method (Fadel, 2010).

Scrum teams are generally small and hold fixed-duration events with the goal of delivering value to their customers. The team is made up of three distinct roles: Scrum Master, Product Owner and Development Team. Each has a specific purpose and is essential for the successful application of Scrum. The Scrum Master's role is to ensure that Scrum is applied correctly, helping to adopt Scrum. The Product Owner is primarily responsible for managing the product backlog and the Development Team is responsible for developing and transforming the product backlog into increments of functionality, creating a ready-to-deliver system. The Backlog is the main artifact of Scrum. It brings together the requirements of the product to be delivered. In other words, it includes all the features, functions, technologies, improvements and corrections that are part of the product to be delivered. It is usually divided into smaller sets that are called the Product Backlog: the entire backlog to be worked on during the project. The Sprint Backlog is only the part of the Backlog selected to be worked on in the Sprint version. The Sprint is an iteration and an event with a fixed duration, which according to the rules must last one month or less and have a goal with a clear objective (Cruz, 2018).

2.6 SOA ARCHITECTURE

Service Oriented Architectures (SOA) is an evolution of distributed computing. This concept was first discussed in the article "Service Oriented Architectures" in April 1996 by researchers Roy Schulte and Yefim Natis of the Gartner Group. SOA is a style of software architecture whose fundamental principle defines that the functionalities implemented by applications must be made available in the form of services (Bigheti, 2020).

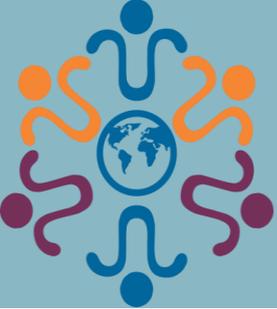

For Xiao *et al.* (2016) these services are connected through a "service bus" (enterprise service bus) that provides interfaces, or contracts, accessible through Web Services or another form of communication between applications. According to Jammes *et al.* (2014) SOA is based on the principles of distributed computing and uses the request/response paradigm to establish communication between client and server systems that implement the services (Figure 1).

Figure 1. SOA Architecture Representation

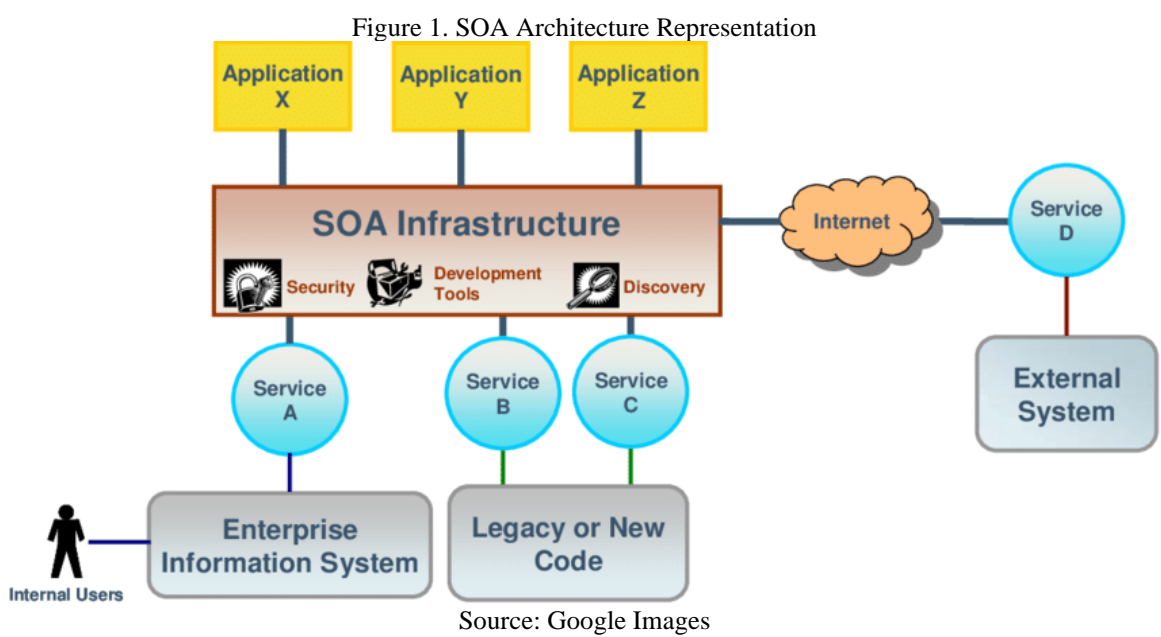

Source: Google Images

SOA is not defined as a technology, methodology or service, but as an architectural concept for systems with the purpose of promoting the integration between functionalities through services (Xiao *et al.,* 2016). This bridge allows exposing the functionalities of applications in standardized and interrelated services. It is conducted through loosely coupled service interfaces, where services do not require technical details of the platform of other services for the exchange of information to be conducted. Many studies, such as those by Xiao *et al.* (2016) and Jammes *et al.* (2014), have focused on the development of SOA with the promise of offering an architecture to support the propagation and use of reusable virtual services.

## 2.7 COW TEAT TEMPERATURE

According to Rodrigues (2018), mastitis is a worrying disease on rural properties due to, once affected, the animal is subject to inflammation or infection of the teats and/or udder, and may or may not present symptoms. Infrared thermography is presented as a non-invasive method that detects physiological functions of body temperature control, in relation to changes in blood flow, which is characterized by an increase in the body's temperature through the surface of the skin. The procedure is safe and quite viable, since there is no need for sedatives, physical restraint, surgical procedures or handling of the animal, although it cannot be used as the only form of diagnosis to date. Furthermore, as described by Rodrigues (2018), when there is a sudden increase in the temperature of the teats or mammary quarters in dairy cows, it may indicate the presence of subclinical mastitis in its early stages and this occurs even before the physical change in the milk or confirmation in tests performed by veterinarians. To determine mastitis in cows through the surface temperature of the teat skin, Table 2 can be used for validation.

Table 2. Cow teat and room temperature indicators

| Corporal superficial (flanco) | Retal superficial | Healthy cow teat | Sick cow teat | healthy mammary glands | Diseased mammary glands |
|---|---|---|---|---|---|
| From 35.8 to 36.5 ℃ | From 36.5 to 37.5 ℃ | From 33 to 34.5 ℃ | 36.5 ℃ | From 32 to 37 ℃ | 38 ℃ |

Source: Rodrigues (2018)

## 3 METHODOLOGY

An agile development methodology was used in the development of this project. The model used during the development of the project was Scrum and its phases are respected in the development of the software.

To develop the system, a database and an application (app) for mobile devices with the Android system in the Java programming language were created. The database was developed in MySQL software, in the cloud service provider SaveInCloud, with access to the Android application via Web Service via HTTP (Hypertext Transfer Protocol).

The application for mobile devices was developed for the Android system and for its communication, a webservice was created to communicate with the server in the cloud using the Apache TomCat WebContainer.

The study was conducted from October to December 2020 on a rural property in the municipality of Alpinópolis/MG with a total area of 4 hectares (20°52'31.8"S 46°18'49.1"W). There were a total of 5 cows, 2 of which were Black and White Holsteins and 3 of which were Girolandos, lactating (but crossbred, without confirmed P.O. – Purebred), aged between 2 and 10 years, with an average of 240 days in lactation and an average production of 11 liters (in total for all cows). The animals were subjected to pasture management, where after the two daily milkings they received 2 kg of concentrated feed.

## 3.1 PROJECT PHASES

The project was divided into two phases with tasks to be performed in each. It was divided into the following phases nominated Application Development and Data Collection, presented below:

a) **application development**: To develop the application, an Android application was developed, which communicates with the database through a Web Service available in the cloud. The activities in this phase include:

1) **web service and database**: a Java application was developed to communicate between the application and the database; the Java programming language was used in the Eclipse IDE to develop the Web Service; a MySQL database was created to persist the data; finally, both services were made available in a cloud service called SaveInCloud;

2) **application**: the application was developed for the Android system using the Android Studio IDE and its respective functionalities were applied in the Java programming language; the application's function is to receive the temperature data from each teat that was collected by the producer during milking; after filling in the data, it is sent to the server for storage;

3) **tests**: the tests to validate the app and its functionalities were performed following the white box standard using the JUnit Framework; to this end, several test scenarios were created and conducted in the application.

b) **data collection**: The data collection schedule was agreed with the farmer responsible for milking the cows. Thus, it was agreed to read the teat temperature at the first milking of the day, around 7 a.m. The digital infrared thermometer model MIDE LX-26E (Visiomed Techonology Co., LTDA, Shenzhen, China) (Visiomed, 2024) shown in Figure 2 was used as the collection tool. The MIDE LX-26E thermometer has a measurement range in body temperature measurement mode: 32°C to 42.9°C and its accuracy is approximately ±0.3°C when measured at a distance of 2 to 5 cm and its calibration guarantee lasts up to 40,000 measurements. All information presented here was taken from the instruction manual provided by the manufacturer. This thermometer was brought approximately 3 to 5 cm from the surface of the cow's teat (as described in the thermometer's instruction manual) and the temperature was obtained and the procedure was repeated for the four teats. The collected data was then manually entered into the MastiteApp application and when the application indicated a change in the temperature reading of one of the cow's teats, the cup test was performed, which consists of taking a small sample of milk in a black-bottomed cup to check for lumps forming at the bottom of the cup.

Figure 2. Infrared Thermometer model MIDE LX-26E

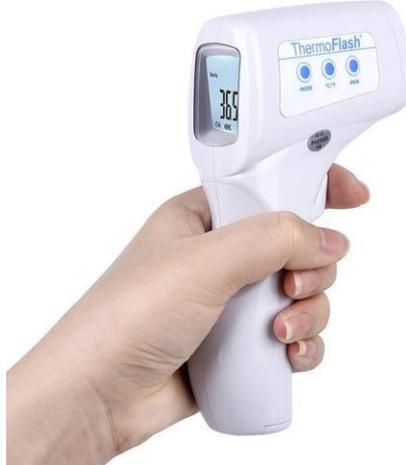

Source: Visiomed (2024)

# 4 RESULTS AND DISCUSSIONS

## 4.1 APPLICATION DEVELOPMENT

### 4.1.1 Development analysis architecture

The project was developed using the agile Scrum methodology, which was used to build the Product Backlog model. During the development of the project, the project screens were prototyped, as shown in Figure 3. This prototyping was necessary both to understand how the application works in relation to its requirements and to validate and approve these requirements according to the perspective of the rural producer:

Figure 3. Prototype of application MastitisApp. a) Home screen; b) record animal data; c) Animal health status; d) Recorded data.

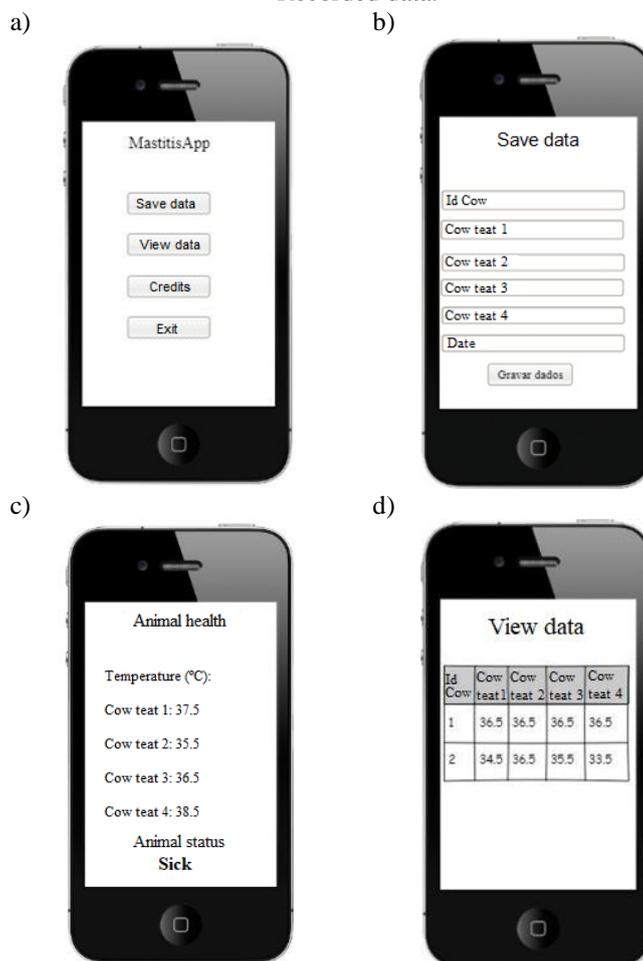

Source: Project authors.

The project was diagrammed before its development based on the prototyping conducted previously. For this purpose, the following diagrams were constructed: Use Cases, Flowchart, SOA Architecture and Table and Relationship Diagram (TRD) of the database. Figure 4 shows the Use Case diagram. This diagram presents two actors, the first being the Farmer, who represents the rural producer, responsible for entering the temperatures read from the infrared thermometer into the application, and the System actor, which represents both the SaveInCloud cloud responsible for storing the collected information and presenting the diagnostic status of the cow teat temperature. In addition, this diagram identified three use cases that were born from the Product Backlog defined in the Scrum methodology.

Figure 4. Use Case Diagram.

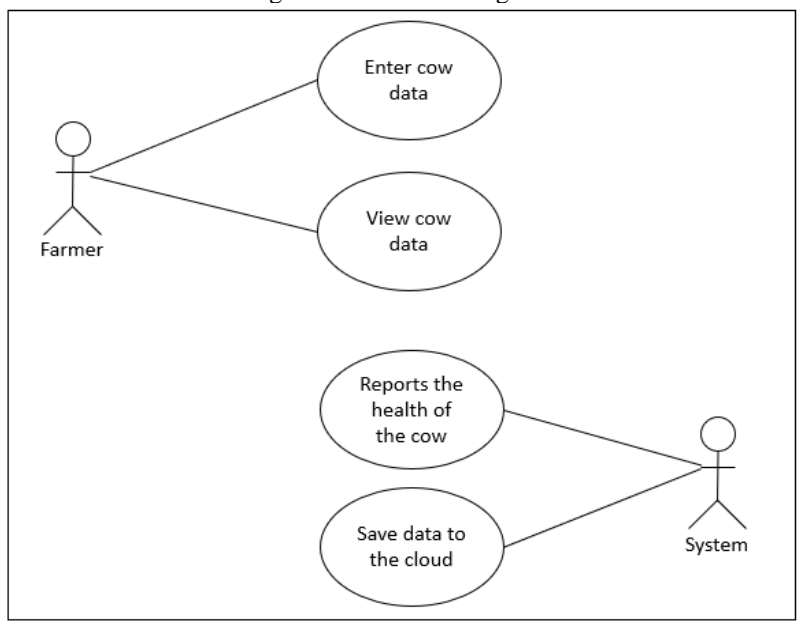

Source: Project authors.

The next diagram chosen for presentation is the Flowchart, as shown in Figure 5. This diagram presents the flow of use of the Mastite App by the rural producer. In this way, the producer reads the teat temperature and, later, enters this temperature into the application and, at the end, the application presents the result of the diagnosis and saves the data in the cloud.

Figure 5. Application execution flowchart.

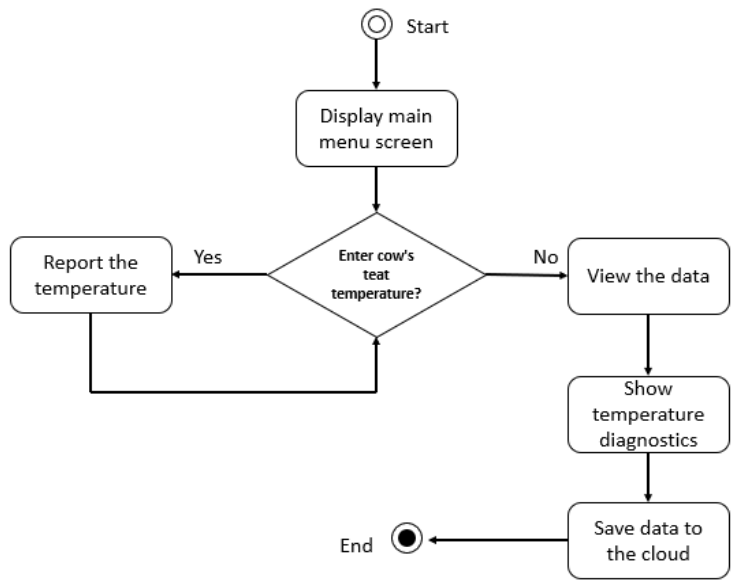

Source: Project authors.

The following is a diagram of the application's cloud architecture, as shown in Figure 6. This architecture shows how data is distributed between the app and the server using the Internet. The MySQL Server database is installed on the server.

Figure 6. SOA Architecture.

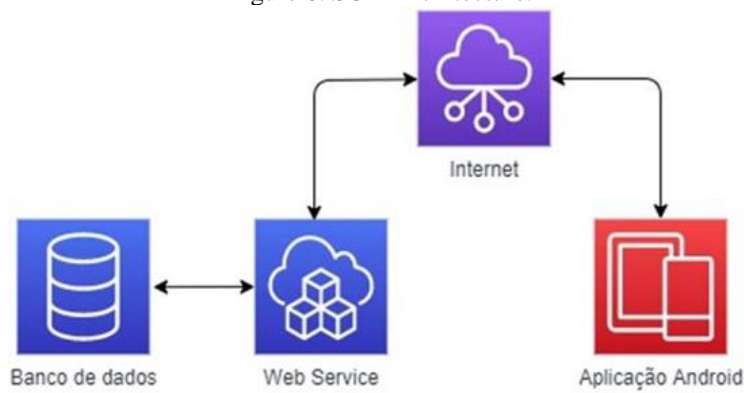

Source: Google Images

Finally, the database model is presented, where only a single table (TRD) was used for data insertion, called Animal Records. This table was configured with seven fields (attributes) that represent: the date of the temperature reading, the temperatures of the four teats, the animal's ID (identifier) and an exclusive field to indicate whether or not those temperatures read indicate the possible presence of subclinical mastitis.

### 4.1.2 Programming architecture

The MastiteApp application was programmed in two IDEs (Interface Development Environments). The first was in Eclipse to build the database manipulation objects and web services as shown in Figure 7.

Figure 7. Class organization in the Eclipse IDE.

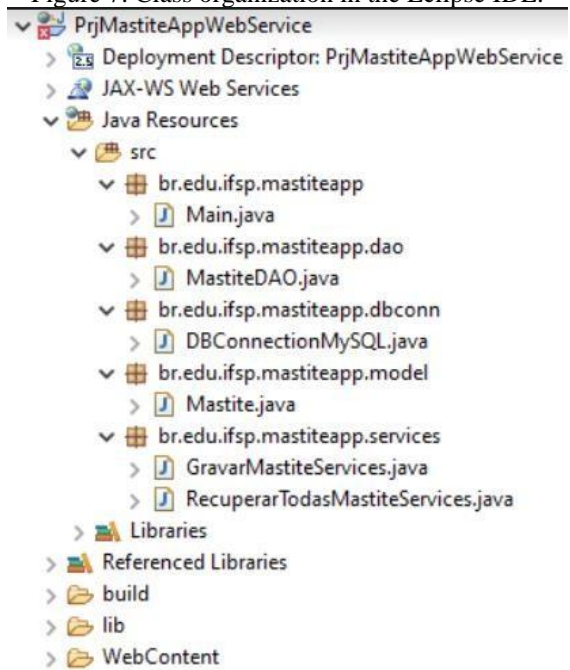

Source: Project authors.

The second was in Android Studio to build the user interface on the Android platform as shown in Figure 8.

Figure 8. Organization of classes in the Android Studio IDE.

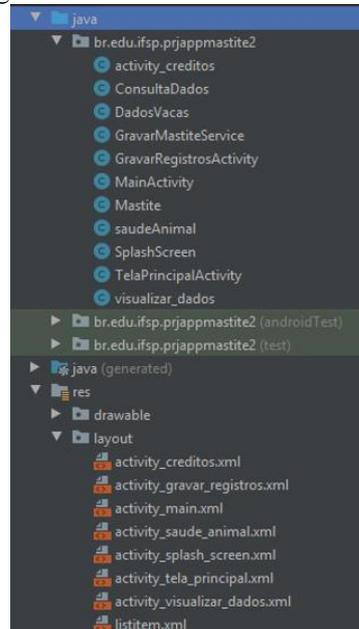

Source: Project authors.

The web services were deployed on the SaveInCloud platform, with support for Java, Apache TomCat 8.5 and MySQL Server, as shown in Figure 9.

Figure 9. Provision of web services.

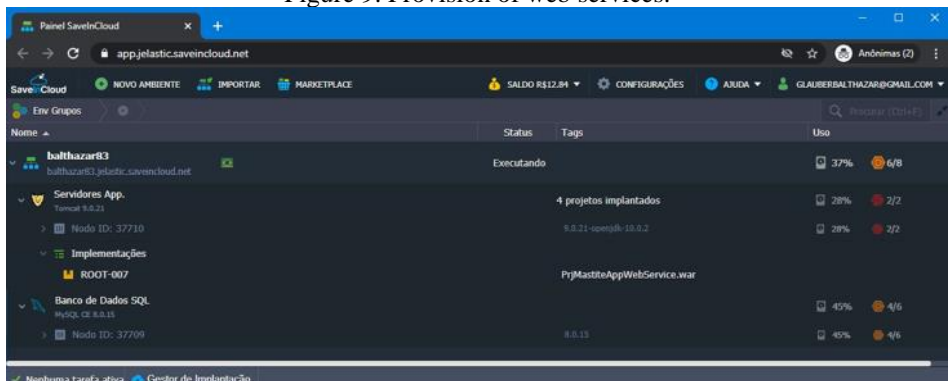

Source: Project authors.

For the data storage interface, a Java code was developed in the Eclipse IDE as shown in Figure 10 to insert the data entered by rural producers through the Android application. This allows the application to consume a web service via an endpoint. This endpoint is described in the code in Figure 10 through the uri variable.

Figure 10. Code for inserting temperatures into the app.

```
18     public GravarMastiteService(Mastite m) { mastite = m; }

19     @Override
20     protected Void doInBackground(Object... objects) {
21         HttpURLConnection urlConnection = null;
22         BufferedReader reader = null;
23
24
25         try {
26             String uri = "http://balthazar81.jelastic.saveincloud.net/ROOT-007/GravarMastiteServices.do?data="+
27                 mastite.getData()+"&is_mastite="+mastite.getMastite()+"&teto1="+mastite.getTeto1()+"&teto2="+mastite.
28                 getTeto2()+"&teto3="+mastite.getTeto3()+"&teto4="+mastite.getTeto4()+"&id_animal="+mastite.getId_animal();
29             URL url = new URL(uri);
30
31             //inicio o processo de conexao
32             urlConnection = (HttpURLConnection) url.openConnection();
33             urlConnection.setRequestMethod("GET");
34             urlConnection.connect();
35
36             InputStream inputStream = urlConnection.getInputStream();
37             reader = new BufferedReader(new InputStreamReader(inputStream));
38             String linha;
39
40             while ((linha = reader.readLine()) != null) {
41                 buffer.append(linha);
42                 buffer.append("\n");
43             }
44         } catch (Exception e) {
45             e.printStackTrace();
```

Source: Project authors.

To insert the data into the database, a web application was developed that receives the parameters of the temperatures of the 4 teats, date, animal identification and presence of mastitis through the reception of an HTTP protocol. Through this application, it is possible to insert the data into the database remotely through a REST (Representational State Transfer) web service via URL (Uniform Resource Locator). As can be seen in the code in Figure 11 through the request service of the HttpServletRequest class (request.getParameter()).

Figure 11. Code for the web service for inserting temperatures.

```
15     public GravarMastiteServices() {
16         super();
17     }
18
19     protected void doGet(HttpServletRequest request, HttpServletResponse response) throws ServletException, IOException {
20         this.executar(request, response);
21     }
22
23     protected void doPost(HttpServletRequest request, HttpServletResponse response) throws ServletException, IOException {
24         this.executar(request, response);
25     }
26
27     protected void executar(HttpServletRequest request, HttpServletResponse response) throws ServletException, IOException {
28         Mastite mastite = new Mastite();
29         mastite.setData(request.getParameter("data"));
30         mastite.setMastite(Integer.parseInt(request.getParameter("is_mastite")));
31         mastite.setTeto1(Float.parseFloat(request.getParameter("teto1")));
32         mastite.setTeto2(Float.parseFloat(request.getParameter("teto2")));
33         mastite.setTeto3(Float.parseFloat(request.getParameter("teto3")));
34         mastite.setTeto4(Float.parseFloat(request.getParameter("teto4")));
35         mastite.setId_animal(Integer.parseInt(request.getParameter("id_animal")));
36
37         MastiteDAO mastiteDAO = new MastiteDAO();
38         mastiteDAO.gravarDados(mastite);
39     }
```

Source: Project authors.

To check whether the cow possibly has mastitis, an interface was developed to check the teat temperature and, if any of them show any temperature change, information is displayed. The code developed to process the data can be seen in Figure 12.

Figure 12. Temperature Check App Code.

```java
public void verificaSaude(Mastite mastite){
    String situacaoAnimal = null;
    TextView textView1 = (TextView) findViewById(R.id.textView15);
    textView1.setText(String.valueOf(mastite.getTeto1()));
    TextView textView2 = (TextView) findViewById(R.id.textView17);
    textView2.setText(String.valueOf(mastite.getTeto2()));
    TextView textView3 = (TextView) findViewById(R.id.textView19);
    textView3.setText(String.valueOf(mastite.getTeto3()));
    TextView textView4 = (TextView) findViewById(R.id.textView21);
    textView4.setText(String.valueOf(mastite.getTeto4()));

    if((mastite.getTeto1()<=34.5&&mastite.getTeto1()>33)||(mastite.getTeto2()<=34.5&&mastite.getTeto2()>33)||
        (mastite.getTeto3()<=34.5&&mastite.getTeto3()>33)||(mastite.getTeto4()<=34.5&&mastite.getTeto4()>33)){
        situacaoAnimal="Saudavel";
    } else if((mastite.getTeto1()<=36.5&&mastite.getTeto1()>34.5)||(mastite.getTeto2()<=36.5&&mastite.getTeto2()>34.5)||
        (mastite.getTeto3()<=36.5&&mastite.getTeto3()>34.5)||(mastite.getTeto4()<=36.5&&mastite.getTeto4()>34.5)){
        situacaoAnimal="Atencao";
    } else if((mastite.getTeto1()>36.5)||(mastite.getTeto3()>36.5)||(mastite.getTeto3()>36.5)||(mastite.getTeto4()>36.5)){
        situacaoAnimal="Doente";
    }
    TextView textView5 = (TextView) findViewById(R.id.textView23);
    textView5.setText(situacaoAnimal);
```

Source: Project authors.

### 4.1.3 Architecture of functional tests

To test the functionality of the services created in the Eclipse Web project, some functional tests were created using the JUnit framework. An example of this functional test is shown in Figure 13.

Figure 13. Test method in Junit.

```java
public class ValidarInsercaoDeDados {

    @Test
    public void GravarInformacoesNoBancoDeDados(){
        MastiteDAO mastiteDAO = new MastiteDAO();
        Mastite mastiteGravacao = new Mastite();
        Mastite mastiteGravado = new Mastite();
        mastiteGravacao.setData("10/10/2020");
        mastiteGravacao.setTeto1(33.54);
        mastiteGravacao.setTeto2(34.54);
        mastiteGravacao.setTeto3(32.23);
        mastiteGravacao.setTeto4(30.66);
        mastiteGravacao.setMastite(1);//conforme + tabela

        mastiteDAO.gravarDados(mastiteGravacao);
        mastiteGravado = mastiteDAO.recuperarUltimoRegistro();
        Assert.assertEquals(mastiteGravado.getData(),"10/10/2020" );
        Assert.assertEquals(mastiteGravado.getTeto1(),33.54,0.001 );
        Assert.assertEquals(mastiteGravado.getTeto2(),34.54,0.001 );
        Assert.assertEquals(mastiteGravado.getTeto3(),32.23,0.001 );
        Assert.assertEquals(mastiteGravado.getTeto4(),30.66,0.001 );
        Assert.assertEquals(mastiteGravado.getMastite(),1,1 );
    }
}
```

Source: Project authors.

In this way, a set of test scenarios were created as shown in Table 3 and then implemented via Java programming.

Table 3. Web Services Test Scenarios.

| Test ID | Description | Expected Result | Result Found |
|---|---|---|---|
| 1 | Record application information in the database | The app is expected to record the following data:<br>Date: 10/10/2020<br>Cow teat1: 33.54<br>Cow teat2: 34.54<br>Cow teat3: 32.23<br>Cow teat4: 38.66<br>Mastitis? Yes<br>Animal ID: 100 | Data has been successfully persisted to the database |
| 2 | Record application information in the database | The application is expected to record the following data:<br>Date: 11/20/2020<br>Cow teat1: 37.28<br>Cow teat2: 39.53<br>Cow teat3: 35.30<br>Cow teat4: 36.51<br>Mastitis? No<br>Animal ID: 246 | Data has been successfully persisted to the database |
| 3 | Record application information in the database | The application is expected to record the following data:<br>Date: 11/20/2020<br>Cow teat1: 38.48<br>Cow teat2: 38.54<br>Cow teat3: 37.54<br>Cow teat4: 36.89<br>Mastitis? No<br>Animal ID: 58 | Data has been successfully persisted to the database |
| 4 | Check last inserted record | The application is expected to return the last data entered as per test number 4. | The data was displayed as entered in the database |

Source: Project authors.

## 4.2 APPLICATION DEVELOPMENT

### 4.2.1 Collection of data

Data collection in the field was conducted using a digital infrared temperature thermometer. Data collection always took place at the first milking of the day around 7 a.m. The temperatures of the four teats of the dairy cows were captured. The data collected using the infrared thermometer were entered into the MastiteApp application. Table 4 shows a summary sample of the temperatures recorded for the five cows with IDs 1, 2, 3, 4, and 5 between October 28 and 31, 2020.

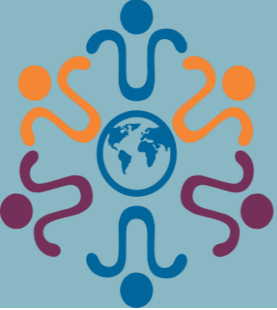

Table 4. Sampling of collected data

| IdCow | Date | Cow teat 1 | Cow teat 2 | Cow teat 3 | Cow teat 4 | Mastitis |
|---|---|---|---|---|---|---|
| 1 | 28/10/2020 | 36.5 | 36.5 | 36.6 | 36.5 | 0 |
| 2 | 28/10/2020 | 36.3 | 36.2 | 36.3 | 36.2 | 0 |
| 3 | 28/10/2020 | 35.9 | 35.7 | 35.8 | 36 | 0 |
| 4 | 28/10/2020 | 35.6 | 35.7 | 35.7 | 35.9 | 0 |
| 5 | 28/10/2020 | 35.9 | 36.0 | 36.1 | 36.0 | 0 |
| 1 | 29/10/2020 | 36.2 | 36.2 | 36.3 | 36.2 | 0 |
| 2 | 29/10/2020 | 36.4 | 36.4 | 36.3 | 35.9 | 0 |
| 3 | 29/10/2020 | 36.2 | 36.1 | 36.0 | 35.9 | 0 |
| 4 | 29/10/2020 | 35.9 | 36.1 | 36.0 | 36.0 | 0 |
| 5 | 29/10/2020 | 35.3 | 35.4 | 35.8 | 35.8 | 0 |
| 1 | 30/10/2020 | 35.8 | 38.7 | 35.8 | 35.8 | 0 |
| 2 | 30/10/2020 | 36.3 | 36.4 | 36.4 | 36.2 | 0 |
| 3 | 30/10/2020 | 35.7 | 35.9 | 35.8 | 35.9 | 0 |
| 4 | 30/10/2020 | 35.7 | 35.8 | 35.8 | 35.7 | 0 |
| 5 | 30/10/2020 | 35.7 | 35.7 | 35.4 | 35.8 | 0 |
| 1 | 31/10/2020 | 36.1 | 36.4 | 36.3 | 36.1 | 0 |
| 2 | 31/10/2020 | 36.2 | 36.2 | 36.4 | 36.2 | 0 |
| 3 | 31/10/2020 | 35.3 | 35.7 | 35.5 | 35.7 | 0 |
| 4 | 31/10/2020 | 36.0 | 36.2 | 35.9 | 36.0 | 0 |
| 5 | 31/10/2020 | 35.4 | 35.8 | 35.8 | 36.1 | 0 |

Source: Project authors.

## 4.2.2 Analysis

The teat condition analysis was performed immediately after data entry into the application. The user then enters the temperature data, clicks the Save button and is automatically directed to a new interface where the result of the diagnosis is presented in relation to the literature that defines the conditions of possible mastitis. This is shown in Figure 14.

Figure 14. Data entry interface and diagnosis of possible mastitis.

a) registration                    b) possible diagnosis

Source: Project authors.

## 5 RESULTS AND DISCUSSIONS

The results and discussions of an article must be presented in a clear and organized manner, based on the data collected and the analyzes carried out during the study. Initially, the results must be presented in an objective and concise way, using tables, graphs and statistics, if applicable, to highlight the main findings. Then, in the discussion section, the results are interpreted in light of existing literature, highlighting similarities, differences and implications for theory and practice.

Furthermore, limitations of the study and possible directions for future research are discussed. It is essential that both the results and the discussion are based on solid evidence and that they contribute significantly to the advancement of knowledge on the topic addressed.

**6 CONCLUSION**

The application proposed in the study was developed and proved to be functional, with a user interface that was easily used by rural producers, attesting that the communication between the application and the WebService is functional and persisting the data in the way it was captured by the producer on his farm. A total of 68 days of temperature analysis were conducted, all collected in the field from the 5 lactating cows, resulting in 160 temperature readings. Based on the information collected in the field, analyses were performed for the presence of subclinical mastitis according to measurements presented in the literature.

In some cases, the temperature was above normal, as presented in the literature, but there were no physical symptoms of mastitis, not even during the other days of measurements. However, in only one case were some lumps present in the mastitis test cup, but the teat temperature was not indicating mastitis as described in the literature.

With these two pieces of information, we can suggest that the number of analyses performed was not sufficient to validate the literature as the only means of predicting the diagnosis of subclinical mastitis, indicating that future work should include an analysis over a longer period of time, aiming to identify confirmed cases of mastitis and comparison with the literature, with the creation of algorithms with the application of artificial intelligence to determine the presence of mastitis.

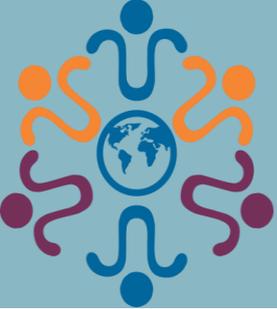

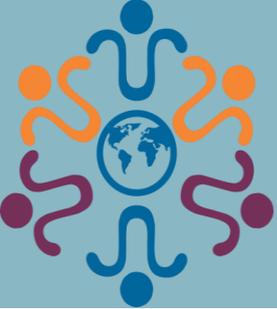

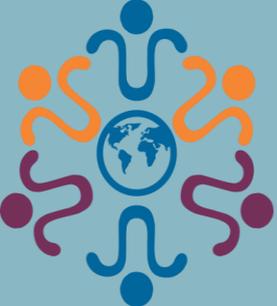